\documentclass[12pt]{amsart}

\usepackage{tikz}
\usepackage{graphicx}
\usepackage{url}

\title[Obtaining Laws of Thermodynamics]{Obtaining Laws of Thermodynamics for Ideal Gases using Elastic Collisions}

\author{Stephen Montgomery-Smith}
\address{Department of Mathematics, University of Missouri, Columbia MO 65211.}
\author{Hannah Morgan}
\address{Department of Computer Science, University of Chicago, Chicago IL 60637}

\begin{document}

\begin{abstract}
The purpose of this note is to see to what extent ideal gas laws can be obtained from simple Newtonian mechanics, specifically elastic collisions.  We present simple one-dimensional situations that seem to validate the laws.  The first section describes a numerical simulation that demonstrates the second law of thermodynamics.  The second section mathematically demonstrates the adiabatic law of expansion of ideal gases.
\end{abstract}

\maketitle

\section{Introduction}

To what extent ideal gas laws can be obtained from simple Newtonian mechanics?  In this note, we present a simple one-dimensional simulation that suggest the answer is ``yes.''.

Similar prior work includes the papers \cite{galant,gall}.  They had a two dimensional model, in which two gases were seperated by a diabatic barrier.  The way energy was communicated between the gases was via the barrier, in which after $n$ collisions, the particles which collided with the barrier in that time were reassigned equal energies, and momenta derived from these energies and their original momenta.  In \cite{hurkala}, this work was used to simulate the Carnot cycle.

This paper gives a slightly different model.  In this paper, the barrier is a massless wall, which is pushed by the particles colliding into it, but the barrier keeps its vertical orientation so that only the horizontal positions of the gas particles matter in the dynamics of the barrier.  An advantage to the approach given in this paper is that the motion is completely Newtonian.  For example, the motion is reversible, thus demonstrating that the second law of thermodynamics is not a hard and fast law, but rather is a trend generated by the pseudo-statistical nature of the deterministic dynamics.

\section{The Second Law of Thermodynamics}

A thermally isolated container has a freely moving piston that splits the container into two parts, the left part and the right part.  The piston is thermally conducting.  The left part is filled with a monatomic gas~A, and the right part of filled with a monatomic gas~B.  Gas~A has an atomic weight $m_1 = 1$, and gas~B has an atomic weight $m_2 = 100$.  Initially, all the molecules have the same average velocity, picked independently from a symmetric random distribution.  The piston is initially placed half way along the container.  We place $N$ molecules of gas~A randomly in the left half of the container, uniformly distributed, and we similarly place $N$ molecules of gas~B in the right half of the container.

The molecules obey Newton's laws of motion, with elastic collisions.  The objective of this note is to describe a numerical experiment which suggests that Newton's laws of motion are all that are needed to predict the ideal gas laws, and that the collisions with the piston should be enough to transfer heat from the hotter gas to the colder gas.  Thus this note is in effect attempting to show that Newtonian mechanics are completely sufficient to explain the second law of thermodynamics.

Molecules from the same gas are assumed sufficiently small so that collisions between them never occur.  Thus the molecules collide either with the walls of the container, or with the piston.  If a molecule collides with a wall, it's velocity is simply reflected.  However the piston has zero mass.  Thus if it is hit by a molecule, it simply travels with that molecule until it hits a molecule of the other gas.  Then the piston acts as a conduit for the two molecules to collide, conserving energy and momentum, as if they had the same $y$-component in space.

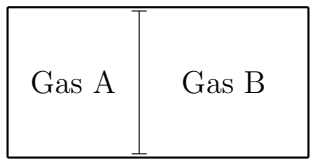
\begin{figure}
\begin{center}
\begin{tikzpicture}
\begin{scope}[scale=0.5]
\draw[thick,-] (-4,2)--(4,2);
\draw[thick,-] (4,2)--(4,-2);
\draw[thick,-] (4,-2)--(-4,-2);
\draw[thick,-] (-4,-2)--(-4,2);
\draw (-0.5,1.9)--(-0.5,-1.9);
\draw (-0.7,1.9)--(-0.3,1.9);
\draw (-0.7,-1.9)--(-0.3,-1.9);
\draw (-2.25,0) node{Gas A};
\draw (1.75,0) node{Gas B};
\end{scope}
\end{tikzpicture}
\caption{Diabatic piston separating two gases in an isolated container.}
\label{piston}
\end{center}
\end{figure}

From the ideal gas law, we have that each container, at least when in a state of quasi-equilibrium, should obey $PV = k NT$, where $P$ is the pressure, $V$ is the volume, $k$ is Boltzmann's constant, $N$ is the number of molecules, and $T$ is the temperature.  Initially the temperature in the left hand part should be $m_2/m_1 = 100$ times cooler than the temperature in the right hand part.  This is because the particles originally have the same average velocities, and hence the right hand side has total kinetic energy 100 times the kinetic energy of the left hand side, and kinetic energy is proportional to temperature.

Initially the pressure in the right hand side will be 100 times the left hand side, and so the piston should move quickly towards the left until the pressures are equalized.  Because this motion is quite fast, one might not expect conditions of quasi-equilibrium to be met, and so there will be some overshoot of this piston due to inertia of the gas molecules.

Next, because of the collisions between the molecules of the different gases, one might expect the kinetic energy to begin to spread evenly between the molecules in the left hand side and the right hand side.  Thus the piston should begin to move slowly towards the middle.

\section*{The Numerical Model}

To compute this numerically, we will completely neglect the vertical motions, and simply compute in one dimension.  We let $N=1000$, and assign to each particle a one dimensional position and momentum.  The only collisions that take place are between molecules and the left or right wall of the container, or between two molecules from the two different gases.  A molecule from gas A will always be to the left of a molecule of gas B.  Molecules from the same gas will simply pass through each other.

The computer program is as follows:
\begin{enumerate}
\item Assign to the $N$ particles of gas~A positions uniformly picked from $[0,0.5)$, and to the $N$ particles of gas~B positions uniformly picked from $(0.5,1]$.
\item Assign to all $2N$ particles velocities uniformly picked from $[-0.5,0.5]$.  (The convention is that velocity is positive if the motion is to the right.)
\item\label{loop} Compute the following collision times:
\begin{enumerate}
\item the time for each particle to collide with the walls $x=0$ or $x=1$;
\item the time for a particle of gas~A to collide with a particle of gas~B.
\end{enumerate}
\item Let $t$ be the smallest of all these collision times.
\item Advance the positions of all the particles to time $t$ using their given velocities.
\item If $t$ represents a collision of a particle with a wall, multiply the velocity of that particle by $-1$.
\item If $t$ represents a collision of a particle of gas~A with velocity $u_1$, and a particle of gas~B with velocity $u_2$, then replace these velocities by $v_1$ and $v_2$, where
\begin{equation*}
v_1 = \frac{(m_1-m_2)u_1 + 2m_2u_2}{m_1 + m_2}, \qquad
v_2 = \frac{2m_1u_1 + (m_2-m_1)u_2}{m_1 + m_2}
\end{equation*}
\item Go back to step~\eqref{loop}.
\end{enumerate}
The program was carefully written so that floating point errors would not cause great error in the calculations.  For example, when two particles collide, they might pass through each other by machine precision.  But then when the collisions are recalculated, they will collide again in a very small amount of time.  Another place where care must be taken is when two collision events take place at exactly the same time.  The collisions were stored in a database, implemented using C++'s standard library {\tt map}.  In this way, after a collision between two particles, or a particle and a wall, the only other collisions that need to be recalculated are those involving these two particles or particle and wall.

\section*{Results}

\subsection*{Volume and temperature}

\begin{figure}
\begin{center}
\includegraphics[scale=0.7]{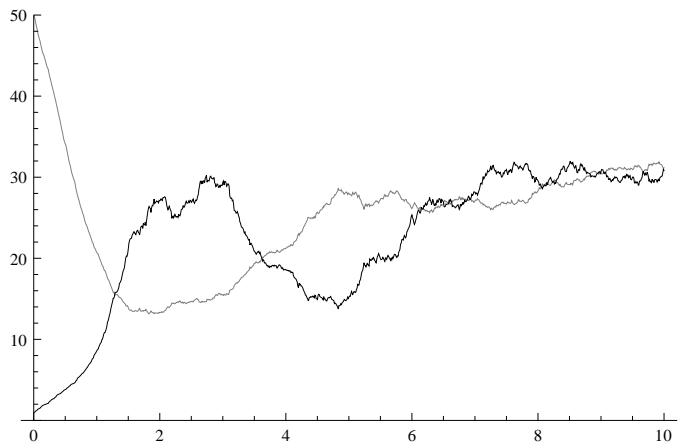}
\caption{Plot of position of piston (gray), and kinetic energy of gas~A over total kinetic energy (black).  THe vertical axis is percentage, and the horizontal axis is time.  This shows the early phase when the piston moves rapidly to the left, and then the system moves towards thermal equilibrium.}
\label{plot1}
\end{center}
\end{figure}

\begin{figure}
\begin{center}
\includegraphics[scale=0.7]{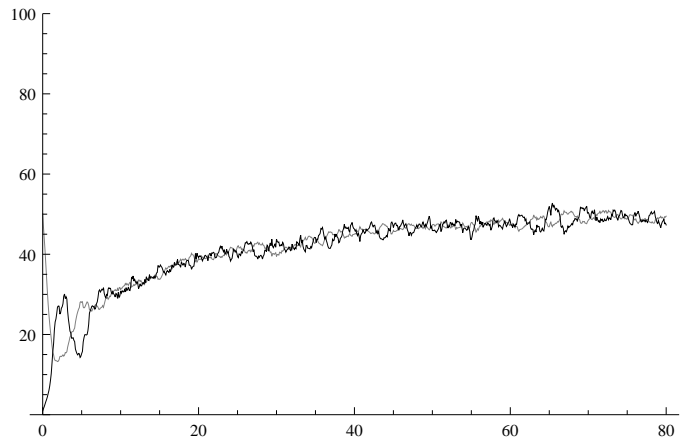}
\caption{This shows the gases achieving thermal equilibrium.}
\label{plot2}
\end{center}
\end{figure}

\begin{figure}
\begin{center}
\includegraphics[scale=0.7]{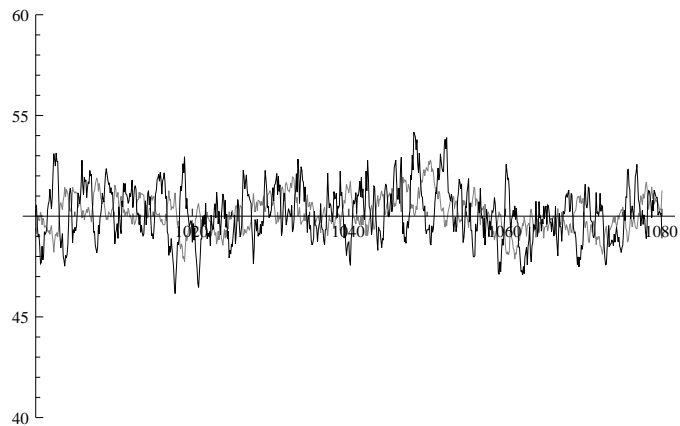}
\caption{Fluctuations long after thermal equilibrium is achieved.}
\label{plot3}
\end{center}
\end{figure}

The results, shown in Figures~\ref{plot1}, \ref{plot2} and~\ref{plot3}, seem to bear out the prediction that Newtonian mechanics is sufficient to obtain the ideal gas laws and the second law of thermodynamics.

\begin{figure}
\begin{center}
\includegraphics[scale=0.7]{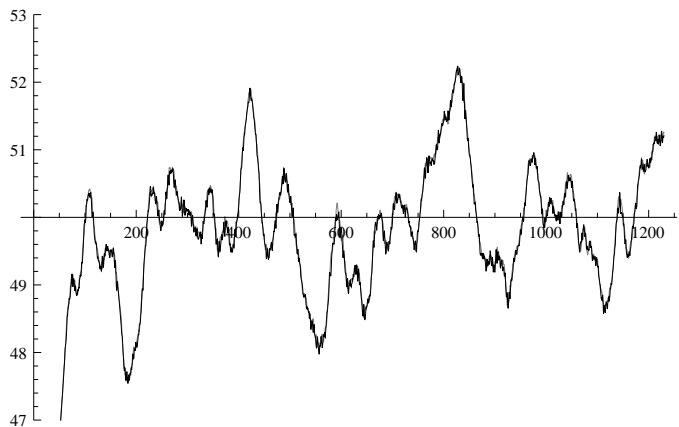}
\caption{Plot of moving averages.  Averaging out the very short time fluctuations shows a tremendous similarity between volume and temperature, validating Charles' Law.  (The position of the piston, shown in gray, is barely discernible behind the kinetic energy of gas A shown in black.)}
\label{plotxxx}
\end{center}
\end{figure}

In Figure~\ref{plotxxx}, we plot the moving averages over a range of 20 time units, showing better than expected agreement with Charles' Law.  The moving averages were calculated using the following formula.
\begin{equation*}
\bar T(t) = \frac1{20}\int_{t-10}^{t+10} T(s) ds
\end{equation*}

\subsection*{Maxwell's distribution of velocities}

For an ideal gas, Maxwell's distribution would predict that the $x$-componant of the velocities of the particles of either gas~A or gas~B should converge to a normal distribution.  We decided to use the Anderson-Darling test described in \cite{stephens,wikipedia}.  We assumed that both the mean and variance of the distributions were unknown.  For each time $t$, we calculated the modified statistic $A^2_m = A^2 (1+4/n-25/n^2)$ as described in Table~1.3 in \cite{stephens}, and as Case~4 in \cite{wikipedia}.  This test assumes that the null-hypothesis is that the distribution is Gaussian, and for a single application of the test, the hypothesis fails with 15\% significance if $A^2_m > 0.571$, and fails with 1\% significance if $A^2_m > 1.092$.

\begin{figure}
\begin{center}
\includegraphics[scale=0.7]{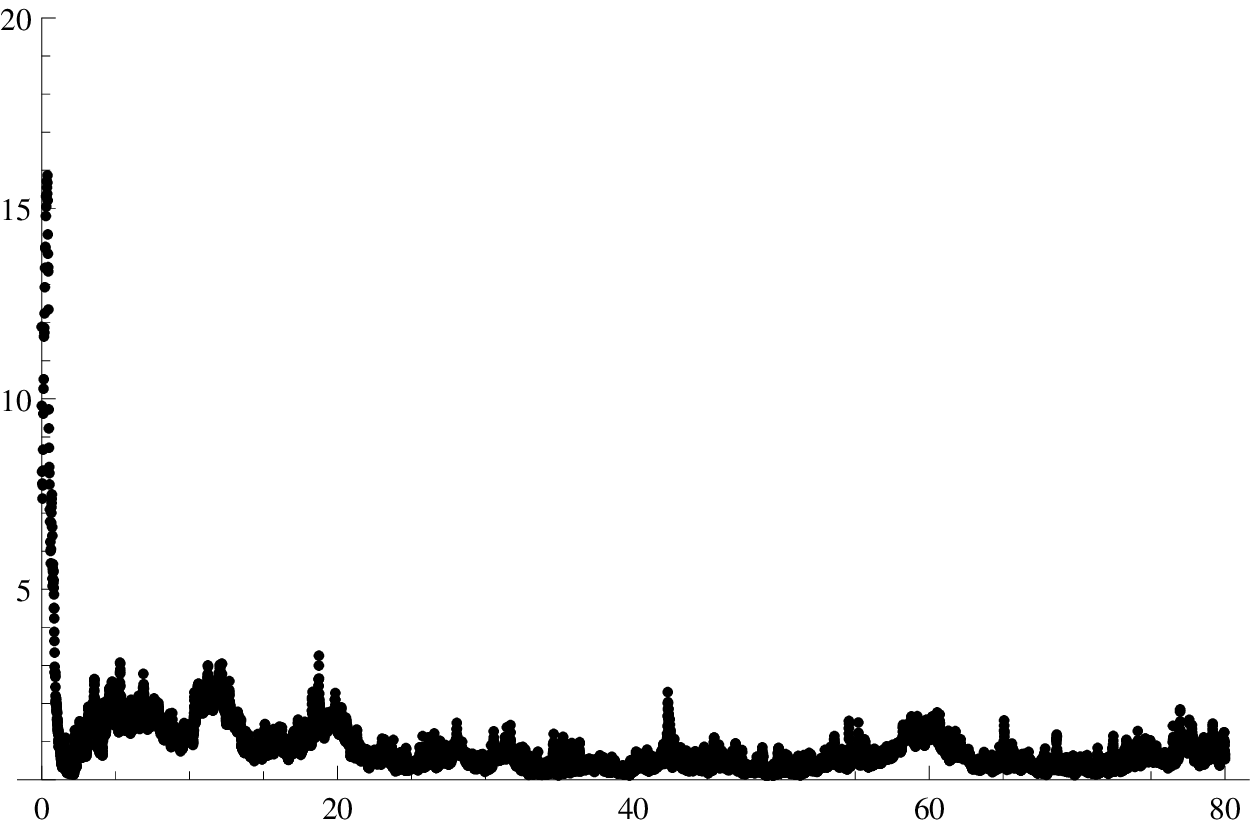}
\includegraphics[scale=0.7]{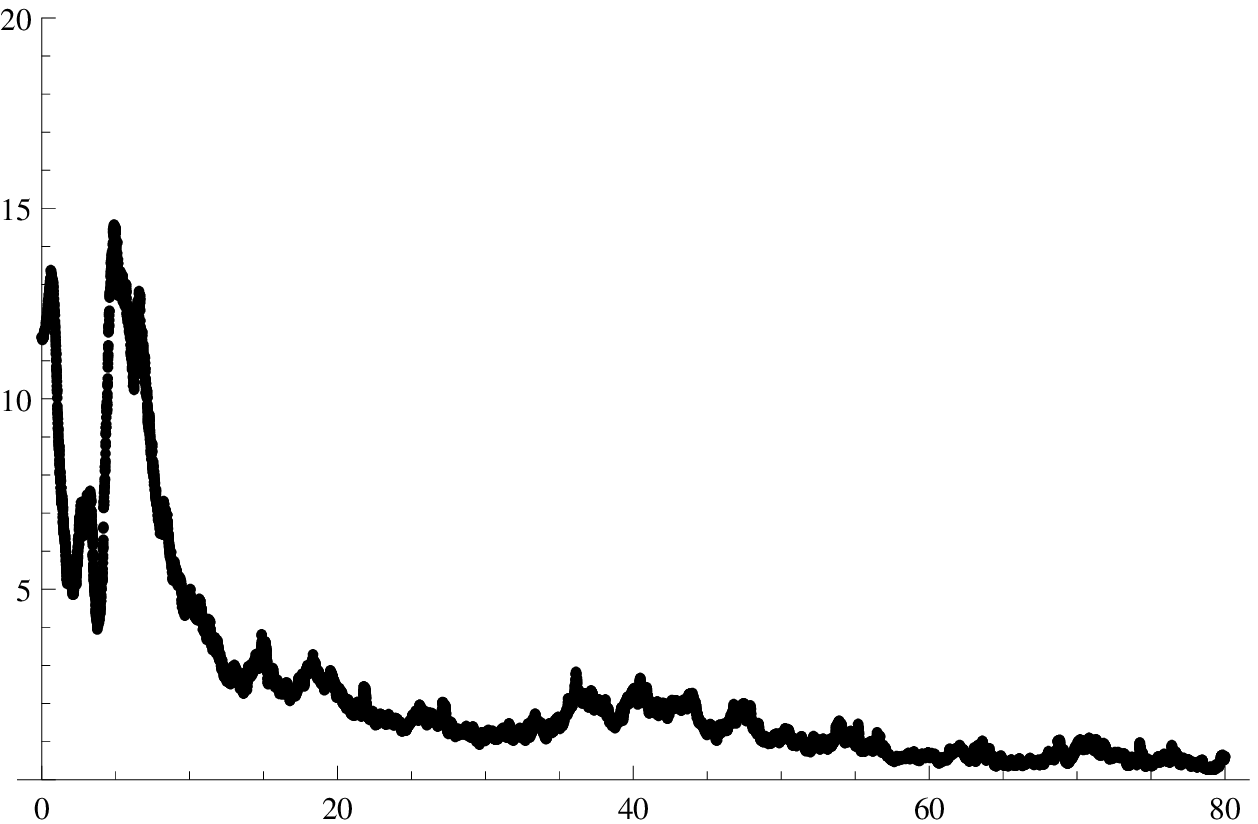}
\caption{This shows the velocities of the gas particles moving towards a Maxwellian distribution, for gas~A and gas~B respectively.  The horizontal axis is time, and vertical axis is the modified $A_m^2$ statistic described in the text.}
\label{plotlr}
\end{center}
\end{figure}

In Figure~\ref{plotlr}, we plot this statistic as a function of time.  This is technically not a proper use of the statistic, as the statistic calculated at one particular time is not going to be statistically independent from the statistic calculated at another particular time.  Nevertheless, the graphs do strongly suggest that both gas~A and gas~B achieve a Maxwellian distribution.  As might be expected, the gas with the heavier molecules, gas~B, takes longer to reach this distribution.

\subsection*{Time reversibility}
Another experiment that was performed was to run the numerics for 1000 time units, and then reverse the velocities, and then run the numerics backwards.  It was found that in most cases that the data after 2000 time units matched the initial data to within about $10^{-11}$.  However in one case the final data was completely different from the initial data.

To help explain why it might be different just one time, the same experiment was performed many times, with only 100 particles and 100 time units.  (The 1000/1000 experiment takes about 3.5 hours.)  In these cases small random perturbations were made to the data at the same time as the velocities were reversed.  By adjusting the random fluctuations to a size of about $10^{-7}$ it was found that the initial data was recovered to within $10^{-5}$ with about the same frequency as the final data being in complete disagreement with the initial data.  In these experiments, a careful log of the order of the collisions was made, and the times when the data agreed closely, the order of collisions in the first 100 time units was almost exactly the reverse of the order of collisions in the second 100 time units (the order differing with at most a few pairs of adjacent collisions reversing their order).  When the final data greatly disagreed with the initial data, at some point in time the collisions became very different.

These all suggest that the apparent statistical nature of equilibrium is NOT due to numerical floating point errors.  It also suggests that if the final data has sensitive dependence on initial conditions, that this only happens when the time elapsed, or the number of particles, is very large.  The effect of ``chaos'' is not primarily responsible for the statistical nature of the equilibrium.

\section{Adiabatic Expansion of Ideal Gases}

In this section, we derive the well known relationship between temperature and volume of an ideal gas when subjected to an adiabatic expansion.  Pressure is a more difficult quantity to understand in our context, so we will leave it out of our discussion.

We have gas particles in a one-dimensional container, whose right hand side is able to move.  Adiabatic means that when the particles collide with the sides, we assume that the collisions are elastic, and the mass of the side walls are effectively infinite.

The volume, $\mathcal V$, is proportional to the distance, $L$, between the left hand side and the right hand side.  The temperature is proportional to the average left-right kinetic energy of the particles, that is, if the $N$ particles have masses $m_n$, and left-right-coordinate of velocity $u_n$, for $1 \le n \le N$, then
$$ \mathcal T\propto \frac1{2N}\sum_{n=1}^N m_n u_n^2 $$
The particles themselves do not interact with each other.  For this reason, without loss of generality, we only need to perform the calculation for one particle.

The only thermodynamic assumption we shall make is that the energy in each particle is spread between the left-right kinetic energy and the other energies equally.  Let us denote by $f$ the ratio of the total energy to the left-right energy of each particle.  Thus if the particle is a three dimensional monotomic gas, then $f = 3$, and if the particle is a three dimensional diatomic gas, then $f = 5$ (because of the extra rotational kinetic energy).  And if the particle is a three dimensional diatomic gas with a bond that is elastic, then $f = 7$ (two extra degrees of freedom for the rotational kinetic energy, one extra degree of freedom for the kinetic energy of the bond oscillating, and one extra degree of freedom for the potential energy of the bond oscillating).

The purpose of this section is to derive the well known adiabatic thermodynamic relation
$$ \mathcal V \mathcal T^{f/2} \text{ is constant} $$

The right hand side wall will move uniformly at a velocity $U$, and the particle will bounce back and forth between the left and right wall.  We assume that the $u_n$ are much larger than $U$, and in the end we shall take the limit as $U/u_n \to 0$.  We also assume that the time for a back and forth bounce is much smaller than the time scale that $U$ might vary.  If $U$ is allowed to vary too quickly, then we can create a kind of ``Maxwell's Daemon" in which we can impart any energy we like to the particle.  The mass of the wall is taken to be $M$, which we will take to be effectively infinite.

We denote the velocity of the $n$th particle before and after it collides with the right wall are $u_n$ and $-v_n$ respectively.  The velocity of the wall after the collision is $V$, and we will soon see that $V = U$.

\begin{figure}
\begin{center}
\begin{tikzpicture}
\begin{scope}[scale=0.5]
\draw[thick,-] (4,1)--(4,-1);
\draw[thick,-] (-4,-1)--(-4,1);
\draw[->] (-3.1,0.5)--node[above]{$u$}(4,0);
\draw[->] (3.8,-0.05)--node[below]{$v$}(-2.9,-0.5);
\draw[->] (3.5,-1.5)-- node[below]{$U$}(4.5,-1.5);
\draw[<->] (-4,1.5)--node[above]{$L$}(4,1.5);
\end{scope}
\end{tikzpicture}
\caption{A particle of gas in an expanding one-dimensional adiabatic container.}
\label{adiabatic}
\end{center}
\end{figure}
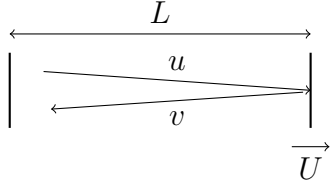

Let us consider one of the particles.  Conservation of momentum implies
$$ m_n u_n + M U = -m_n v_n + M V $$
Conservation of energy implies
$$ f m_n u_n^2 + M U^2 = f m_n v_n^2 + M V^2 $$
Solving these equations we obtain
\begin{equation*}
v_n = - \frac{(m_n-f M)u_n + 2MU}{f M + m_n}, \qquad
V = \frac{2f m_n u_n + (f M-m_n)U}{f M + m_n}
\end{equation*}
Now letting $M \to \infty$, we obtain
$$ v_n = u_n - \frac2f U, \qquad V = U $$
The time $\delta t$ required for the particle to bounce from the right hand wall, to the left hand wall, and back to the right hand wall is $\delta t = 2L/u_n$.  In that time, the length of the wall changes $\delta L = U \delta t$, and the change in absolute value of the velocity of the particle is $\delta u_n = - \frac2f U$.  Then we obtain
$$ \delta L = \frac{2UL}{u_n} = - \frac{f L\delta u_n}{u_n} $$
Now let $U/u_n \to 0$.  Then we obtain
\begin{gather*}
\qquad \frac{dL}{du_n} = -f \frac L{u_n} \\
\Rightarrow \qquad d(\log L) = - f d(\log u_n) \\
\Rightarrow \qquad L u_n^f \text{ is constant} \\
\Rightarrow \qquad L^{2/f} u_n^2 \text{ is constant}
\end{gather*}
Averaging over $n$ we obtain
$$ \mathcal V^{2/f} \mathcal T \text{ is constant} $$

\section{Conclusions}

The authors feel that there are many extensions to these kinds of numerical simulations that could provide more illustrations of certain properties of ideal gases.  For example, in the first section we could perform the experiment assuming that a constant force is applied to the piston, or we could assume the piston has non-zero mass.  Potentially there is a large number of variations that could be performed.

We also feel that the scenario described in the first section could be a place to look for plausible theorems that could be rigorously proved.  In particular, it seems to the authors that equilibrium is achieved rather quickly, and probably much more quickly than could be proved if we assumed certain ergodicity assumptions on the phase space of solutions.  Perhaps equilibrium theorems could be proved without ergodicity assumptions.  And of course any equilibrium results must only show that equilibrium is achieved most of the time, because as our backwards in time calculations show, in this discrete setting the second law of thermodynamics is more of a trend than a law.

Finally, from an education point of view, we feel that these simulations provide a very concrete way to understand thermodynamic laws.

\end{document}